\documentstyle[12pt]{article}

\begin{document}
\begin{center}
\large \bf{CONVECTION AND HEAT TRANSFER IN LIQUID UNDER LOW GRAVITY CONDITIONS
AND THERMOCAPILLARY EFFECTS.}\\
Gadiyak G.\ V., Cheblakova E.\ A. \\
\normalsize \rm{
Institute of Computational Technologies,\\
Siberian Division of the Russian Academy of Sciences,  \\
Lavrentyev av. 6, Novosibirsk, 630090, Russia\\}
E-mail address: lena@net.ict.nsc.ru\\
Fax number: (3832) 341342\\
Fone number: (3832) 342280\\
\end{center}
\normalsize \rm
\section*
{Abstract}

The two-dimensional flow of viscous incompressible liquid in a square cavity 
with a free boundary and differentially heated vertical sides is considered
in the present work. The influence of gravitational and thermocapillary
convection on temperature and velocity fields is studied in large range of
dimensionless parameters and similarity criteria using equations in a 
Boussinesq approximation.
Limiting cases of dimensionless parameters are analyzed numerically.

\section*
{Introduction}

It is known that in the nonuniformly heated fluid motion appears. Without
free boundaries it appears due to thermal (gravitational)
convection. In the nonuniformly heated liquid with a free boundary there arises
thermocapillary convection \cite{1}, \cite{2}. Under the low gravity condition
it is necessary to take into account both of these processes. It is interesting to
analyze the influence of these factors on heat fluxes and velocity fields
formation under the non-gravity condition and with an increase in gravity.

We will consider a plane stationary convectional system with a side heating.
Such a formulation of the problem is realized in practice, for example, in
crystal growth equipment, various energy plants \cite{2}, \cite{3}. That is why
the knowledge of flow and heat fluxes structure in the large 
dimensionless parameters range (Rayleigh $Ra$, Prandtl $Pr$ and Marangoni
$Ma$ numbers) is of scientific and practical interest.

\section*
{Model}

The stream function - vorticity $(\psi-\omega)$ formulation of the problem was 
used. For plane geometry initial non-dimensional stationary convection 
equations in a Boussinesq approximation under the non-gravity condition 
are the following \cite{4}:
\begin{eqnarray}
\frac{\partial }{\partial x}(u\omega )
+\frac{\partial }{\partial y}(v\omega ) &=& \frac{Pr}{Ma}\bigtriangledown ^2 
\omega,\label{e:1.1} \\
\bigtriangledown ^2 \psi &=& \omega ,\label{e:2.1} \\
\frac{\partial }{\partial x}(uT)+
\frac{\partial }{\partial y}(vT) &=& \frac{1}{Ma}\bigtriangledown ^2 T. \label{e:3.1}
\end{eqnarray}
where
$u = \frac{\partial \psi}{\partial y},\ 
v = -\frac{\partial \psi}{\partial x},$ 
$\ \omega = \frac{\partial u}{\partial y}-\frac{\partial v}{\partial x}.$ 

The problem is characterized by the following parameters: the Marangoni number
$Ma=\frac{\bigtriangleup T \sigma_T L}{\mu a}$ and the Prandtl number
$Pr=\frac{\mu}{\rho a}.$ Here $\bigtriangleup T = T_H-T_C,$ $T_H$ is the 
hot wall temperature, $T_C$ is the cold wall temperature,
$\sigma _T$ denotes the temperature coefficient of surface tension,
$L$ stands for the characteristic length (the side of a square cavity), $\mu$ is the dynamic
viscosity, $\rho$ designates the density, $a$ is the thermal diffusivity.

Calculations with various Prandtl and Marangoni numbers were carried out.
Two limiting cases were also considered: $Ma \rightarrow 0$ and
$Pr \rightarrow \infty.$
If the Marangoni number tends to zero (i.e. surface tension force is equal to
zero) then the set of Eqs. (\ref{e:1.1} -- \ref{e:3.1}) is modified as 
follows:
\begin{eqnarray}
\bigtriangledown ^2 \omega &=& 0,\label{e:1.3} \\
\bigtriangledown ^2 \psi &=& \omega ,\label{e:2.3} \\
\bigtriangledown ^2 T &=& 0. \label{e:3.3}
\end{eqnarray}
If the Prandtl number tends to infinity (that is the case of the strongly viscous
fluid) then the set of Eqs. (\ref{e:1.1} -- \ref{e:3.1}) is developed as
follows:
\begin{eqnarray}
\bigtriangledown ^2 \omega &=& 0,\label{e:1.4} \\
\bigtriangledown ^2 \psi &=& \omega ,\label{e:2.4} \\
\frac{\partial }{\partial x}(uT)+
\frac{\partial }{\partial y}(vT) &=& \frac{1}{Ma}\bigtriangledown ^2 T. \label{e:3.4}
\end{eqnarray}

If the force of gravity $g$ is not equal to $0,$ initial stationary 
non-dimensional equations of
thermal convection in a Boussinesq approximation in the uniform gravitational field
in $(\psi-\omega)$ variables are then \cite{4}:
\begin{eqnarray}
\frac{\partial }{\partial x}(u\omega )
+\frac{\partial }{\partial y}(v\omega ) &=& \frac{Pr}{Ma}\bigtriangledown ^2 
\omega-\frac{Ra\cdot Pr}{Ma^2}\frac{\partial T}{\partial x} ,\label{e:1.5} \\
\bigtriangledown ^2 \psi &=& \omega ,\label{e:2.5} \\
\frac{\partial }{\partial x}(uT)+
\frac{\partial }{\partial y}(vT) &=& \frac{1}{Ma}\bigtriangledown ^2 T. \label{e:3.5}
\end{eqnarray}
Here $Ra=\frac{\beta \rho g L^3(T_H-T_C)}{a\mu}$ is the Rayleigh number,
$\beta=-\frac{1}{\rho}\frac{\partial \rho}{\partial T}$ denotes the volumetric
coefficient of thermal expansion.

In the present work the following boundary conditions are considered.
The vertical sides are at temperatures $T=T_H=0{,}5$ ("hot" wall) and 
$T=T_C=-0{,}5$ ("cold" wall).
The lower horizontal wall and the free surface $y = 1$ are insulated. Both velocity 
components are zero on the walls. On the free surface the component
$\mu\frac{\partial u}{\partial y}$ of the viscous tension tensor is equal to
the tangential force acting on the surface
$-\sigma_T \frac{\partial T}{\partial x},$ and vertical velocity $v$ is equal 
to zero. Thus the boundary conditions for the considered equations can be written in
the following way:\\
$\psi = \frac{\partial \psi}{\partial x} = 0,\qquad T = T_H = 0{,}5 \qquad $ at $x = 0,$ \\
$\psi = \frac{\partial \psi}{\partial x} = 0,\qquad T = T_C = -0{,}5 \qquad $ at $x = 1,$ \\
$\psi = \frac{\partial \psi}{\partial y} = 0,\qquad 
\frac{\partial T}{\partial y} = 0 \qquad $ at $y = 0,$\\
$\psi = 0,\qquad \frac{\partial ^2 \psi}{\partial y ^2} = 
-\frac{\partial T}{\partial x}, \qquad
\frac{\partial T}{\partial y} = 0 \qquad $ at $y = 1.$

The case of the surface tension force equal to zero was also considered.
This corresponds to the system (\ref{e:1.3} -- \ref{e:3.3}) with the
boundary condition
$\frac{\partial u}{\partial y}|_{y=1} = 0:$\\
$\psi = 0,\qquad \frac{\partial ^2 \psi}{\partial y ^2} = 0
\qquad $ at $y = 1.$

\section*
   {Method of solution}

Formulation of the problem (\ref{e:1.1} -- \ref{e:3.1}), as many 
other problems of viscous incompressible fluid in $(\psi, \omega)$ variables
has the following difficulty. Boundary conditions on the walls are determined
only for the stream function, and not for the vorticity, which is defined only
inside the domain according to the Eq. (\ref{e:2.1}). To overcome this
difficulty various approaches are used, for example, approximate boundary 
conditions for vorticity. In the present work we use the Toma condition for 
$\omega$ on the wall \cite{5}:
\begin{eqnarray}
\omega _{k} = \frac{2(\psi _{k+1}-\psi _k)}{\Delta h^2}+O(\Delta h).
\end{eqnarray}
Here $\Delta h$ denotes the mesh size, $\psi_k$ is the value of the stream function
in the boundary node $k$, $\psi_{k+1}$ is is the value of $\psi$ in the node
$k+1$ nearest to the wall.

Calculations were also performed with the second order boundary condition for 
$\omega,$ namely, Woods condition. Computational results with these different
conditions are almost the same. However, the use of the approximate boundary
conditions for vorticity on the wall for the Eqs. (\ref{e:1.1} -- \ref{e:3.1}) 
at high Marangoni numbers and on fine meshes leads to
the considerable slowing-down of the convergence. That is why we also used 
the boundary conditions calculation method \cite{6}, which allowed us to improve
the convergence in $2-4$ times, and in some cases on the order in comparison with
the use of the Toma formula.

Idea of this method is to determine the boundary condition for vorticity inside
the main domain, where $\omega$ is defined according to (\ref{e:2.1}).
Equation for $\omega$ (\ref{e:1.1}) is solved in the auxiliary domain. The 
solid boundaries of this domain are displaced on the mesh size into the cavity 
from the solid 
boundaries of the main domain. The free surface $y = 1$ is common for these 
domains. On the free surface vorticity is determined as follows:\\
$\omega |_{y=1}=\frac{\partial u}{\partial y}|_{y=1}=
-\frac{\partial T}{\partial x}|_{y=1}.$\\
Stream function and temperature equations are solved in the main domain. There 
are two boundary conditions for the stream function. The condition
$\psi _0 = 0$ is used to solve equation for $\psi.$ 
The obtained stream function field does not satisfy the gradient condition
$(\frac{\partial \psi }{\partial n})|_{x=0,x=1,y=0} = 0$ yet. Therefore, the
values of $\psi$ on the boundary of the auxiliary domain are corrected with the help
of the difference analog of this condition \cite{6}. Using a three-point 
approximation of the second order for the derivative $(\frac{\partial \psi }
{\partial n})|_{x=0,x=1,y=0}$ we will obtain:\\
at $x=0 \quad \psi _{2 j}=\frac{1}{4} \bar{\psi}_{3 j}, \ j=2,\ldots,m-1;$\\
at $x=1 \quad \psi _{n-1 j}=\frac{1}{4} \bar{\psi}_{n-2 j}, \ 
j=2,\ldots,m-1;$\\
at $y=0 \quad \psi _{i 2}=\frac{1}{4} \bar{\psi}_{i 3}, \  i=2,\ldots,n-1.$

When approximating Eqs. (\ref{e:1.1}, \ref{e:3.1}) and
(\ref{e:1.5}, \ref{e:3.5}) for temperature and vorticity exponential fitting
discretization (or scheme of integral identities) was used \cite{7}, \cite{8}.
It allowed us to obtain a higher
precision in comparison with the usual approximations. As a result, a five-point
algebraic system was carried out. Equations for $\omega $ and $T$ do not
satisfy the diagonal dominance condition. It is known that without this
condition, many effective methods fail to converge or converge very slowly.
In the present work modification of Buleev's method \cite{9} and a splitting
method \cite{10} are used. They enable to find a solution of the system without
a diagonal dominance property. Buleev's method modification converges faster
than the splitting method when the stream function and the temperature
equations are solved. But for the vorticity equation with high Marangoni numbers 
($Ma>10^2$) Buleev's method does not converge to the necessary precision
(namely, $1.e-7$). Thus the equation for $\omega$ was solved by splitting which 
enables to obtain the prescribed precision.

To improve convergence the damping operation for vorticity was used. It is 
determined by the following recurrent relation:
\begin{eqnarray*}
\omega ^{n+1}_d = \theta \omega ^{n+1} + (1-\theta )\omega ^n_d,
\end{eqnarray*}
where $\theta$ stands for the damping parameter,
$\omega ^{n}_d$ is the damped value of vorticity from the $n$ iteration,
$\omega ^{n+1}$ denotes the value of $\omega$ from the $n+1$ iteration.

During numerical experiments for a $61\times 61$ mesh the damping parameter 
$\theta $ was about $0{,}002.$

\section*
{Main results and their analysis}

We use rectangular non-uniform thickening to the boundary of the $21\times 21,$ 
$41\times 41$ and $61\times 61$ area grids. Calculations were performed at
the Marangoni numbers from $10^{-3}$ to $10^4,$ the Prandtl numbers from $1$ to $100$
and the Rayleigh numbers from $0$ to $10^6.$ Two limiting cases: $Ma \rightarrow 0$
and $Pr \rightarrow \infty$ were also considered. Results are presented in 
Figures (1-10) and in Tables I and II on the $61\times 61$ mesh.
Results at the Marangoni number $Ma = 10^4$ are not very reliable as the
calculating scheme converges badly at this value of dimensionless parameter.

Figure 1 shows the influence of surface tension force on the temperature field
under the non-gravity condition ($Ra=0$). If the Marangoni number $Ma= 0,$ i.e.
the surface tension force is equal to zero (Fig.1a), there is no convection, and
the fluid rests. With an increase in the Marangoni number the convective
mixing enhances, and at $Ma = 10^4$ the flow becomes vortical. One can see
from Figures 1b, 1c, 1d, that the greater $Ma,$ the stronger the
contour maps deflect and are pressed to the hot and cold walls.
Near these walls temperature boundary layers appear. The boundary layer near
the hot wall is wider than near the cold. One characteristic property of the 
picture is the condensation of contour maps to the upper right corner of the
cavity, to the region near the free surface and the cold wall. In the center of
the domain a vortex is formed. Contour maps on Figure 1 correspond to the
Prandtl number $Pr = 1.$ Figure 2 shows temperature contour maps at $Pr = 
\infty$ (that is the case of the strongly viscous fluid, Eqs.
(\ref{e:1.4} -- \ref{e:3.4})). Behavior near the free surface
$y = 1$ is different for the corresponding contour maps of Figures 1 and 2.
In the case of the strongly viscous fluid contour maps come normally to the free
surface and remain undeflected near it longer than in the case of $Pr = 1.$ 
The explanation is that it is more difficult for the surface tension force to 
mix more viscous fluid. From the comparison of contour maps at $Ma = 10^4$ 
(Figures 1d and 2c) one can see that at $Pr = 1$ a vortex region is larger
and extend almost over the entire domain. At $Pr = \infty$ the vortex is
situated closer to the free surface, convective flow in the entire domain is 
weaker. Thus the more viscosity of the fluid, the weaker convection there at the
same surface tension force. 

Figure 3 presents the influence of the gravity ($Ra$) on the temperature field.
Analyzing the results, we can say that with the increase in Rayleigh number the
whole picture of the flow becomes more complex, convection mixing of the fluid
intensifies, the vortical structure of the stream and narrow boundary
layers with the sharp temperature difference across the boundary layer near 
the hot and cold walls appear.
Comparison of corresponding contour maps of Figures 1 and 3 shows that the surface
tension force tends to form a vortex in the center of the domain and to 
condense the contour maps to the upper part of the cold wall; whereas the force of
gravity tends to form a vortex closer to the boundary of the cavity, 
remaining its center without a vortex, and to condense the contour
maps to the vertical walls. At $Ra=10^6$ two narrow boundary layers near the
vertical sides of the width $\approx 0{,}07$ each, with horizontal temperature
gradient are generated. In the remaining part of the calculation domain of the 
width $\approx 0{,}85$
the contour maps are located at a greater distance and the temperature gradient points 
vertically upwards. Such behavior of the temperature contour maps influences
the stream function field. One can see from Figure 4d for the $\psi $ contour 
maps ($Ra=10^6$) that near the vertical walls contour maps condense, and two 
narrow boundary layers with the sharp difference of stream function values
across the boundary layer are formed. In the remaining part of the calculation 
domain the contour maps are located at a greater distance. At $Ra=10^5$ (Fig.4c) the
secondary flows are formed in the center of the domain.

Table I shows an increase in the maximum value of velocity module
$\sqrt{u^2+v^2}$ and the change in its location with an increase in the
Rayleigh number. At $Ra\leq 10^5$ this maximum is situated on the free surface,
and the greater $Ra$, the closer the maximum to the hot wall. At $Ra=10^6$ the maximum
value of $\sqrt{u^2+v^2}$ is located close to the center of the hot wall.

\begin{table}
\begin{center}
\begin{tabular}{|c|c|c|c|}   \hline
$Ra$ & $max|v|$ & $x$ & $y$  \\ \hline 
$10^2$ & 11,871  & 0,6776 & 1   \\ 
$10^3$ & 13,639  & 0,6776 & 1  \\ 
$10^4$ & 26,626  & 0,5    & 1 \\ 
$10^5$ & 73,827  & 0,3225 & 1   \\ 
$10^6$ & 192,88  & 0,0541 & 0,5969   \\ \hline
\end{tabular}
\end{center}
\caption{Dependence of the maximum value of velocity module and its location
on the Rayleigh number.}
\end{table}

Dependence of the maximum value location of $\sqrt{u^2+v^2}$ on the
Marangoni number was built (Fig.5). For all considered $Ma$ numbers the
maximum value of velocity module is situated on the free surface. With an
increase in $Ma$ it moves to the upper right corner, the common point of the
cold wall and the free surface. For a more viscous fluid this maximum is located
closer to the hot wall for the same Marangoni numbers. It is evident that the
surface tension force tends to move the vortex to the upper boundary of the
cold wall, and it is more difficult to mix more viscous fluid. In the range of
$Ma$ from $10^3$ to $10^4$ some oscillation of the curve can be seen (Fig.5). 
It is connected with the more complex structure of the flow at the high
Marangoni numbers. A high precision is necessary
for the calculation of the flow at the Marangoni numbers $Ma\geq 10^4.$
Such precision could not be obtained with the help of the
proposed scheme. 

Figure 6 presents the contour maps of stream function at $Pr=1,$ $Ra=0$ for various
Marangoni numbers. With an increase in $Ma$ the contour maps begin to come
closer to the upper corners, and in the lower corners the secondary flows 
appear. With the increase in the
Marangoni number the inner vortex domain increases, and the contour
maps value module corresponding to this vortex decreases. The contour maps
condense near the free surface.

Figure 7 shows $\psi $ contour maps for the strongly viscous fluid at different
Marangoni numbers. One can see from a comparison of Figures 6 and 7
that in more viscous fluid the convective flow in the entire calculation domain 
is weaker, 
vortex is located closer to the free surface and vortex's region is less;
there are no secondary flows in the lower corners of the domain.

Let us compare Figures 6a ($Pr = 1, \ Ma = 10^2, \ Ra =
0$) and 4a ($Pr = 1, \ Ma = 10^2, \ Ra = 10^3$). With the appearance of the
thermal gravitational convection the contour maps shape
changes. The lower parts of the contour maps are located closer to 
the bottom of the domain, and contour maps tend to the lower corners.
It is the result of the gravity force influence. The greater
$Ra,$ the stronger the influence of the gravity force in comparison with the
surface tension force on the character of the flow. Thus at $Ra = 10^3, \ 
Ma = 10^2$ (Fig.4а) the contour maps also tend to the upper right corner, i.e.
the thermocapillary convection considerably influences the contour maps
shape. With the further increase in $Ra$ the gravitational convection becomes dominant.
Beginning from $Ra=10^5$ the secondary flows are formed.
One of them tends to the upper
left corner of the cavity, to the hot wall, another - to the lower right corner, 
to the cold wall (Fig.4b, 4c, 4d). 

Calculations of the convection problem under the non-gravity condition with the
surface tension force equal to zero (Eqs. (\ref{e:1.3} -- \ref{e:3.3}) 
with the zero boundary condition $\frac{\partial u}{\partial y}|_{y=1} = 0)$ were
made. Here no forces operate on the fluid, there is no convection, and all
profiles of stream function, velocities and vorticity at various sections of 
the calculation domain are almost zero. The temperature contour maps in this
case are straight lines (Fig.1a). It means that the heat is transferred only
by heat conduction.

Calculations of the convection problem without gravitation were
also carried out with the following parameters: $Ma = 1;\ 0{,}1;\ 0{,}01;\ 
0{,}001;\ Pr = 1.$ With the decrease in the Marangoni number the stream
function and velocity profiles tend monotonously to the zero values, to
the profiles of the limiting case $Ma = 0.$ The temperature contour maps also
tend monotonously to the contour maps of the case $Ma=0$ with the decrease in
$Ma.$

Another limiting case: $Pr \rightarrow \infty$ for various Marangoni numbers
(Eqs. (\ref{e:1.4} -- \ref{e:3.4})) under the non-gravity
condition was also considered. Computational results show that for every
Marangoni number between
$0$ and $10$ the flow does not depend on the Prandtl number (at
$Pr$ between $1$ and $\infty$). At
$Ma = 10^2$ and $Pr$ from $10$ to $\infty$ the corresponding profiles of the 
flow agree 
(Fig.2a (temperature contour maps), Fig.7a (contour maps of $\psi $)), 
but differ from the flow profiles at the Prandtl number $Pr = 1$ (Fig.1b, 6a). 
A similar
situation takes place at $Ma = 10^3$ (Fig.1c, 6b; 2b, 7b). At $Ma = 10^4$ and 
$Pr = 100, \infty$ the profiles are close to each other and differ from the
flow profiles at $Pr = 10.$

For the heat balance control the average Nusselt numbers on the hot and cold walls
and on the vertical mid-plane of the cavity were calculated. The local Nusselt 
number $Nu(x,y)$ at the point $(x,y)$ and the average Nusselt number
$Nu_{x_0}$ on the section $x = x_0$ are determined as follows:

\begin{eqnarray*}
Nu(x,y) &=& Ma\cdot uT-\frac{\partial T}{\partial x},\\
Nu_{x_0} &=& \int_0^1Nu(x_0,y)dy.
\end{eqnarray*}

Table II shows the computational results of the convection problem in the cavity
with a free surface and side heating under the non-gravity condition at 
$Pr = 1$ and various Marangoni numbers. Here $u(1/2,1)$ is the value of $u$
velocity at the mid-point of the free surface, $Nu_0,\ Nu_{1/2},\ Nu_1$ designate
the average Nuselt numbers on the sections $x=0, \ x=1/2, \ x=1$ respectively.
The obtained results were compared with \cite{11}. In the work \cite{11} a high
accuracy scheme was used to solve Eqs. (\ref{e:1.1} -- \ref{e:3.1}). 
One can see from the Table that the present work method allows us to
obtain with acceptable precision the integral characteristics of heat-removing
($Nu_0,\ Nu_{1/2},\ Nu_1$). However, the local characteristics ($Nu(0,1),\ 
Nu(1,1)$) are determined with essential errors, particularly at $Ma=10^4.$

\begin{table}
\begin{center}
\begin{tabular}{|c|c|c|c|c|c|c|c|}   \hline
$Ma$ & grid & u(1/2,1) & $Nu_0$ & $Nu_{1/2}$ & $Nu_1$ & $Nu(0,1)$ & $Nu(1,1)$ \\ \hline 
$10^2$ & $21\times 21$ & 1,0326(-1) & 1,1322 & 1,0999 & 1,0396 & 0,7972 & 1,7816 \\
 & $41\times 41$ & 1,0711(-1) & 1,1010 & 1,0977 & 1,0867 & 0,7402 & 1,9874 \\ 
 & $61\times 61$ & 1,0671(-1) & 1,0945 & 1,0953 & 1,0924 & 0,7313 & 1,9905 \\ \hline
[11] & $61\times 61$ &  1,0869(-1) & 1,0962 & 1,0962 & 1,0962 & 0,7301 & 2,028 \\ \hline
$10^3$ & $21\times 21$ & 5,2221(-2) & 2,1800 & 1,9976 & 1,6900 & 1,3444 & 7,3224 \\
 & $41\times 41$ & 5,2032(-2) & 1,9858 & 1,9742 & 1,8983 & 1,0268 & 10,6884 \\
 & $61\times 61$ & 5,1404(-2) & 1,9437 & 1,9528 & 1,9297 & 0,9707 & 11,438 \\ \hline
[11] & $61\times 61$ &  5,0018(-2) & 1,9258 & 1,9258 & 1,9258 & 0,9550 & 11,75 \\ \hline
$10^4$ & $21\times 21$ & 2,7263(-2) & 5,4250 & 3,9362 & 2,6086 & 5,1195 & 12,6693 \\
 & $41\times 41$ & 3,1541(-2) & 4,5622 & 4,3744 & 4,1286 & 3,1227 & 38,5290 \\
 & $61\times 61$ & 3,1562(-2) & 4,4459 & 4,4467 & 4,3524 & 2,5091 & 60,471 \\ \hline
[11] & $61\times 61$ &  3,0381(-2) & 4,3621 & 4,3621 & 4,3654 & 2,2334 & 77,09 \\ \hline
 \end{tabular}
\end{center}
\caption{Computational results at Pr=1.}
\end{table}

Dependencies of the average Nusselt numbers on the Prandtl, Marangoni and 
Rayleigh numbers (Fig.8, 9 and 10 respectively) were also calculated. Fig.8
shows that the greater the Marangoni number for the same Prandtl
number (i.e. the stronger thermocapillary convection), the stronger 
heat-removing. 
At $Ma$ between $0$ and $10^2$ the heat-removing coefficient
does not change with the increase in $Pr.$ At $Ma=10^3;\ 10^4$ and from $Pr=10$
this coefficient takes some constant value, and every Marangoni number has 
its own value. Besides, at $Ma$ between $10^3$ and $10^4$ an abrupt increase  
in the Nusselt
number is observed, especially for the case $Pr=1.$ 
Probably, the reason for it lies in the appearance of a more complex structure
of the flow and in the loss of it the stability \cite{11}, which lead to
the sharp increase in heat-removing. Such an abrupt increase in heat-removing is also
observed in Fig.9, which presents dependence of $Nu$ on $Ma$ for various
Prandtl numbers. The Nusselt number increases with the growth of the Rayleigh
number, which characterizes the intensity of the gravitational convection
(Fig.10). At $Ra$ between $10^4$ and $10^6$ abrupt increase in the Nusselt 
number also 
occurs because of the appearance of a more complex flow structure. Figures 9,
10 present functions that are the parameter approximations of dependences of
the Nusselt
number on the hot wall on the Rayleigh number at $Ra$ between 
$10^5$ and $10^6$ and on the Marangoni number at $Ma$ between $10^3$ 
and $10^4.$ In figures these functions are showed by triangles
$(\bigtriangleup \bigtriangleup \bigtriangleup).$ 
They are the linear combinations of functions with the fitting coefficients
as follows:\\
$g(Ma)=1{,}481+7{,}15\cdot10^{-4}\cdot Ma-3{,}277\cdot10^{-8}\cdot 
Ma^2$ - in the case of the Nusselt number on the Marangoni number dependence;\\
$g(Ra)=4{,}122+9{,}507\cdot10^{-6}\cdot Ra-4{,}296\cdot10^{-12}
\cdot Ra^2$  - in the case of the Nusselt number on the Rayleigh number 
dependence.\\


\section*
{Conclusions}

1. The influence of gravitational and thermocapillary convection on the 
temperature and velocity fields has been studied. Computational results show that
at $Ma\leq 10^2$ and
beginning from $Ra=10^4$ the gravitational convection becomes dominant.\\
2. With the increase in the gravity force the whole picture of the flow
becomes more complex, convection mixing of the fluid
intensifies, narrow boundary layers with the sharp temperature and velocity 
difference across the boundary layer close to the hot and cold walls appear.
Beginning from $Ra=10^5$ the secondary flows formation occurs.\\
3. In a more viscous fluid the convective flow in the entire calculation domain 
is weaker, 
vortexes are located closer to the free surface and vortexes region is less.\\
4. Two limiting cases: $Ma \rightarrow 0$ and $Pr \rightarrow \infty$ 
without gravitation were considered. Computational results show that
with the decrease in the Marangoni number the flow profiles tend monotonously
to the flow profiles of the case $Ma=0.$ Ranges of the dimensionless parameters
where the flow does not change were also determined.\\
5. The parameter approximations of dependences of the Nusselt
number on the hot wall on the Rayleigh number at $Ra$ between 
$10^5$ and $10^6$ and on the Marangoni number at $Ma$ between $10^3$ 
and $10^4$ were built.

\section*
{Acknowledgements}

The authors thank Berdnikov V.\ S., Gaponov V.\ A.,
Cheblakov G.\ B., Dvurechenskii A.\ V. for helpful discussions and encouragement. 
This work was supported by Russian
Fund of Fundamental Researches under Grant N 96-01-00137.         

\newpage
\addcontentsline{toc}{section}{\protect\numberline{6}{Список литературы}}

\newpage
\large \bf {FIGURES}\\
\rm \normalsize
Fig.1. Contour maps of temperature $T:$ $Pr=1,$ $Ra=0,$ \\
a - $Ma=0,$ b - $Ma=10^2,$ c - $Ma=10^3,$ d - $Ma=10^4.$\\
Fig.2. Contour maps of temperature $T:$ $Pr=\infty,$ $Ra=0,$ \\
a - $Ma=10^2,$ b - $Ma=10^3,$ c - $Ma=10^4.$\\
Fig.3. Contour maps of temperature $T:$ $Pr=1,$ $Ma=10^2,$ \\
a - $Ra=10^3,$ b - $Ra=10^4,$ c - $Ra=10^5,$ d - $Ra=10^6.$\\
Fig.4. Contour maps of stream function $\psi :$ $Pr=1,$ $Ma=10^2,$ \\
a - $Ra=10^3,$ b - $Ra=10^4,$ c - $Ra=10^5,$ d - $Ra=10^6.$\\
Fig.5. Maximum value location of the velocity module versus the
Marangoni number (--- $Pr=1,\ - - Pr=\infty$).\\
Fig.6. Contour maps of stream function $\psi :$ $Pr=1,$ $Ra=0,$ \\
a - $Ma=10^2,$ b - $Ma=10^3,$ c - $Ma=10^4.$\\
Fig.7. Contour maps of stream function $\psi :$ $Pr=\infty,$ $Ra=0,$ \\
a - $Ma=10^2,$ b - $Ma=10^3,$ c - $Ma=10^4.$\\
Fig.8. Average Nusselt number on the hot wall versus the Prandtl
number \\
(1 - $Ma=0,$ 2 - $Ma=10^2,$ 3 - $Ma=10^3,$ 4 - $Ma=10^4$).\\
Fig.9. Average Nusselt number on the hot wall versus the
Marangoni number \\
($g(Ma)=1{,}481+7{,}15\cdot10^{-4}\cdot Ma-3{,}277\cdot10^{-8}\cdot Ma^2$) \\
--- $Pr=1,\ - - Pr=\infty,$ $\bigtriangleup \bigtriangleup \  g(Ma).$ \\
Fig.10. Average Nusselt numbers on the hot and cold walls versus
the Rayleigh number 
($g(Ra)=4{,}122+9{,}507\cdot 10^{-6}\cdot Ra-4{,}296\cdot10^{-12}\cdot Ra^2$) \\
--- $Nu0,\ - - Nu1,$ $\bigtriangleup \bigtriangleup \  g(Ra).$
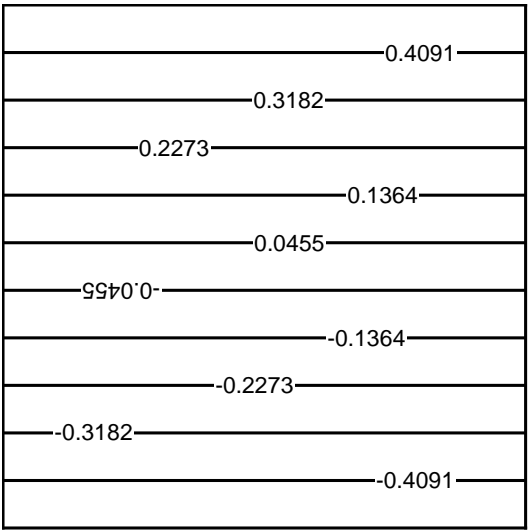
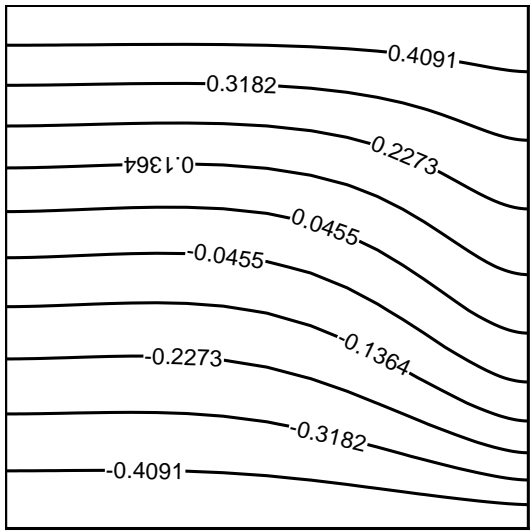
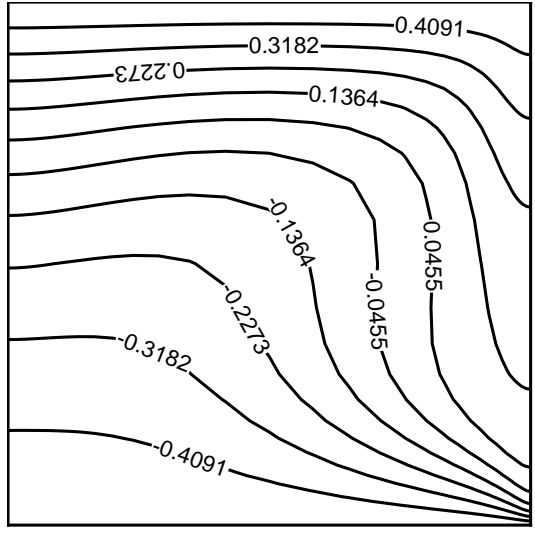
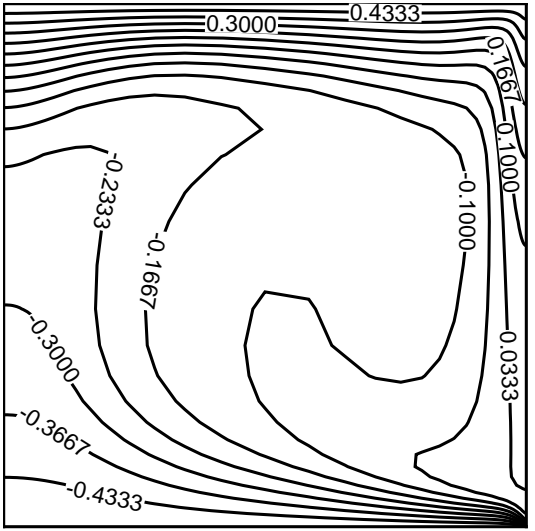
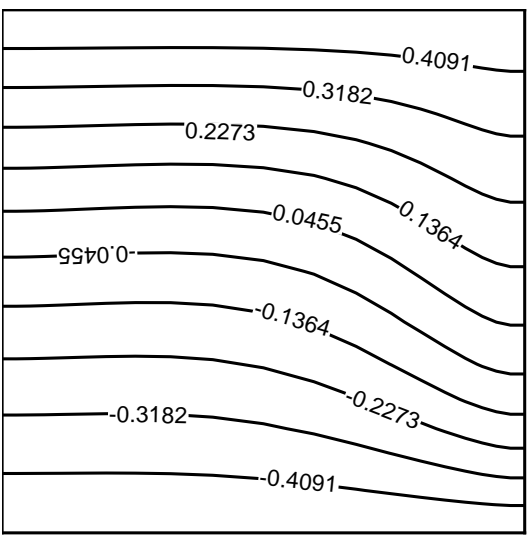
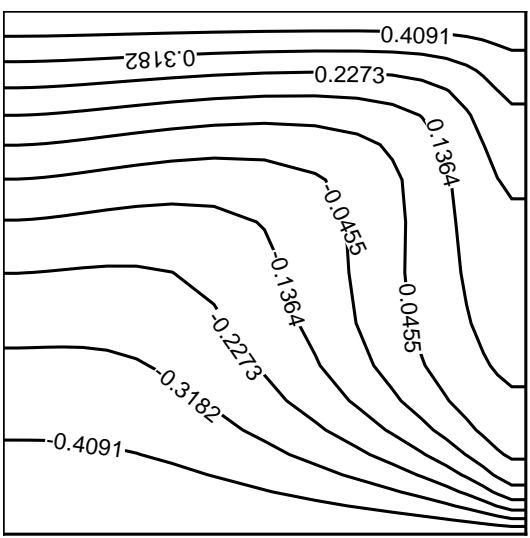
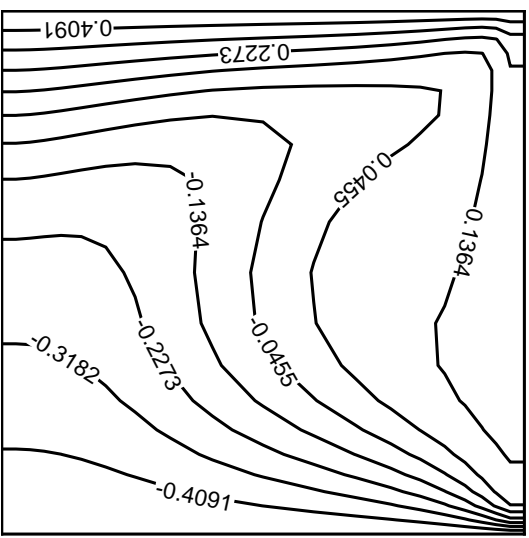
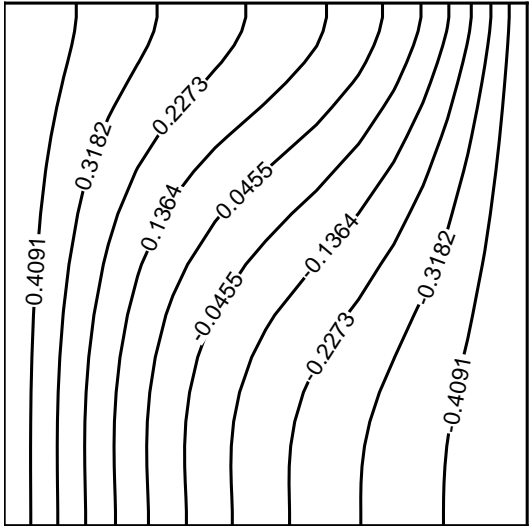
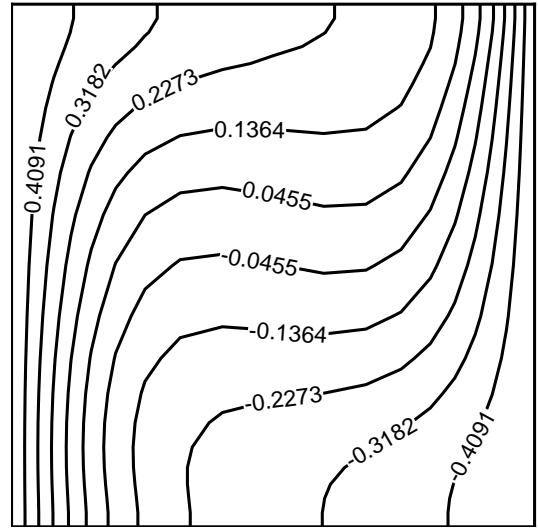
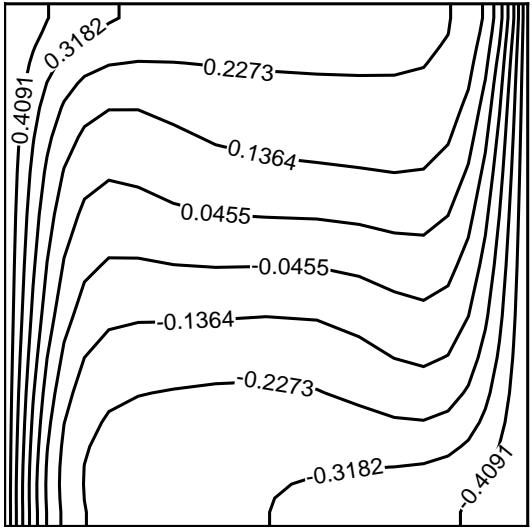
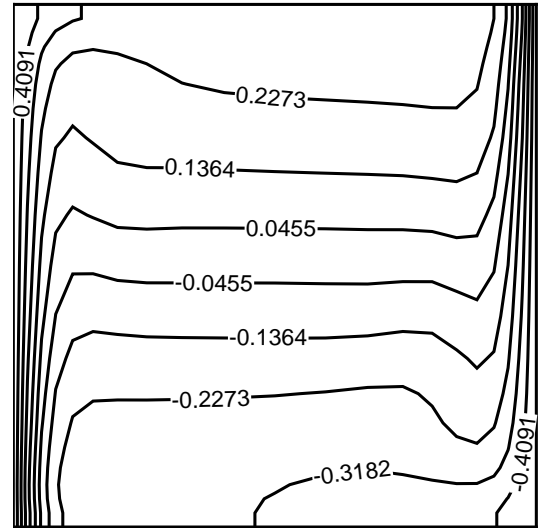
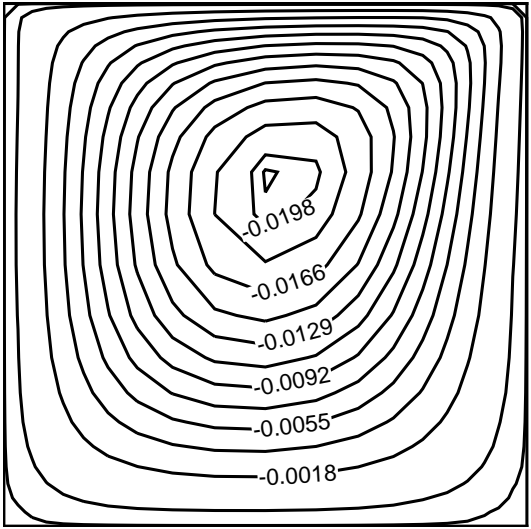
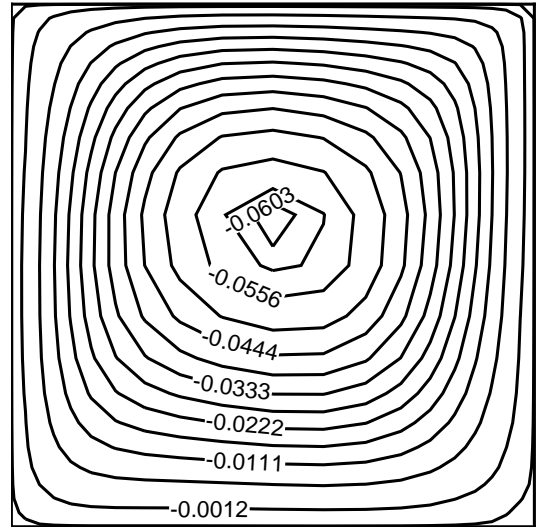
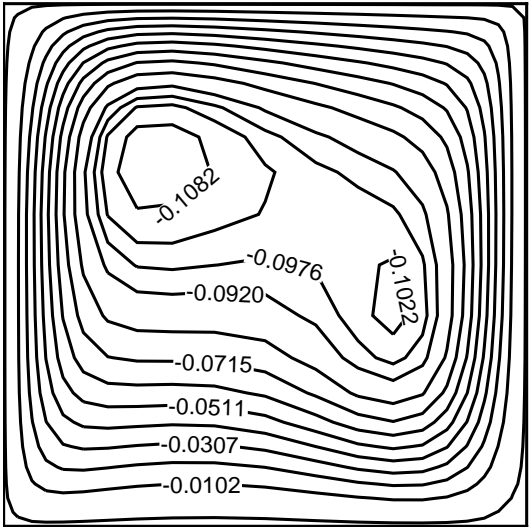
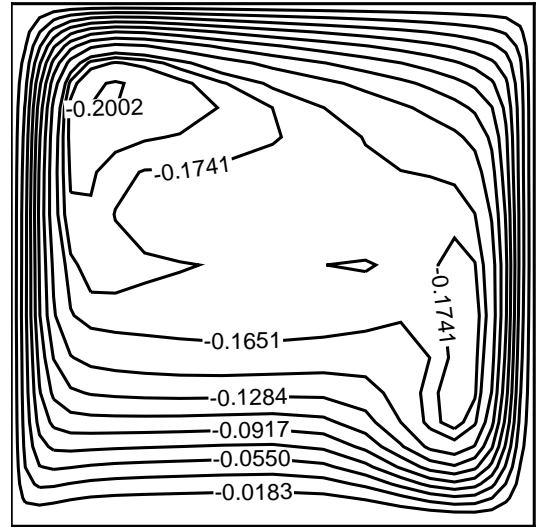
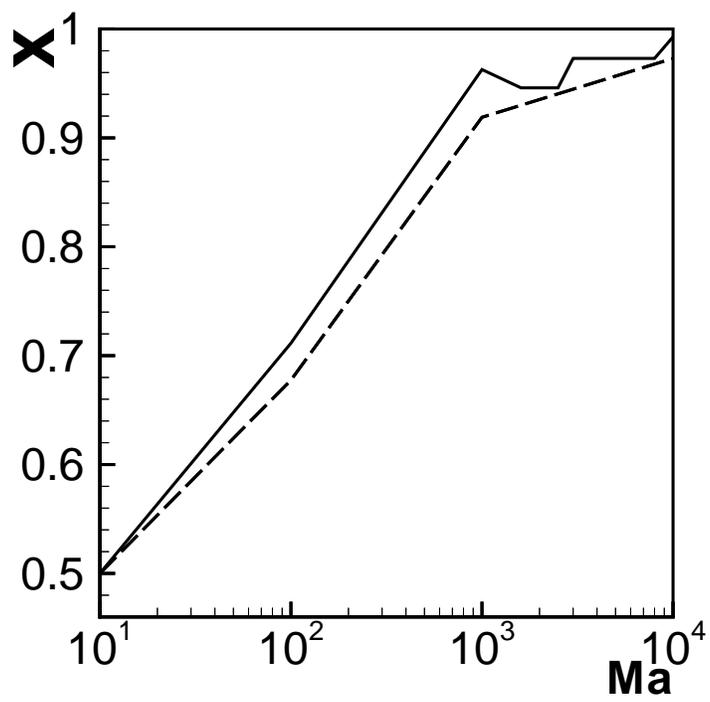
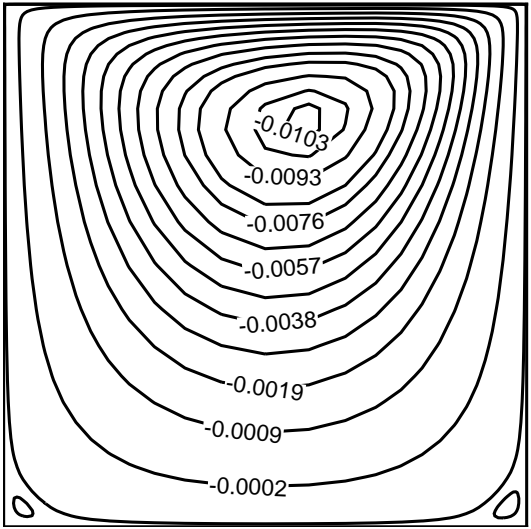
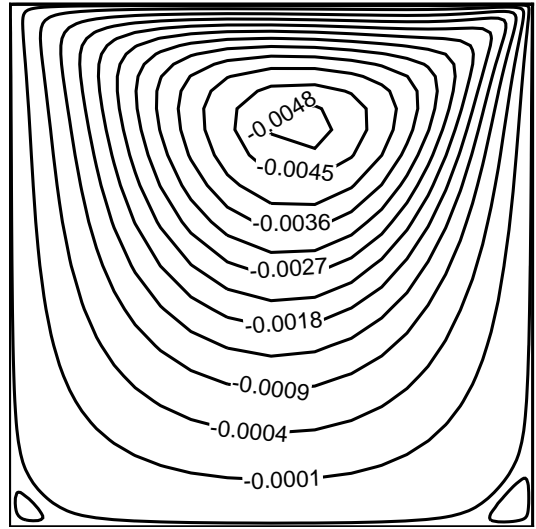
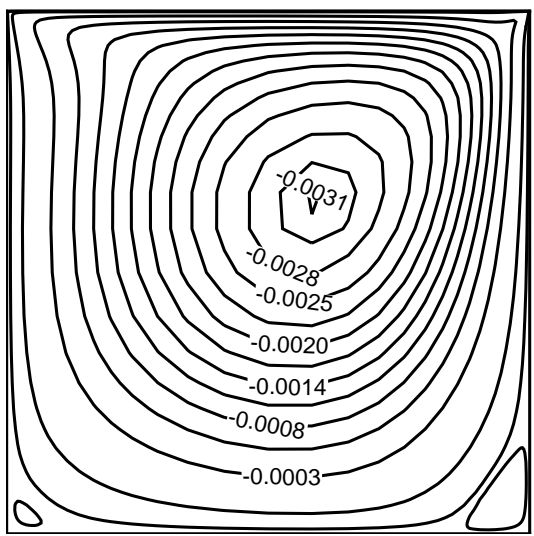
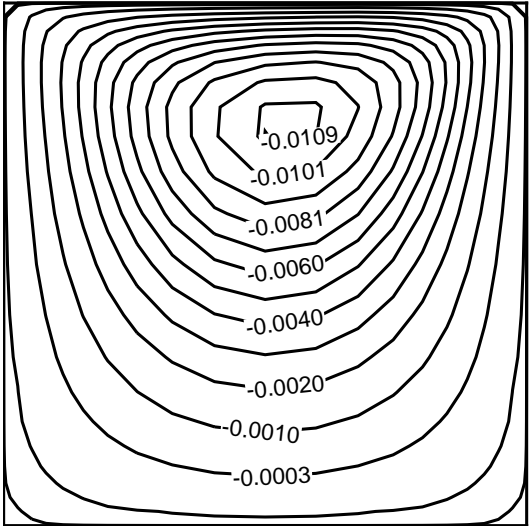
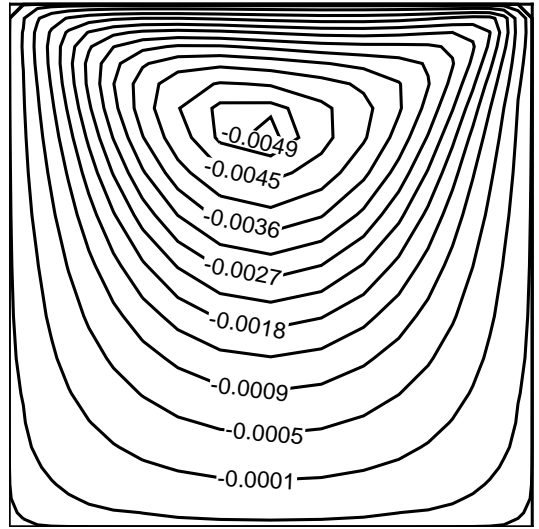
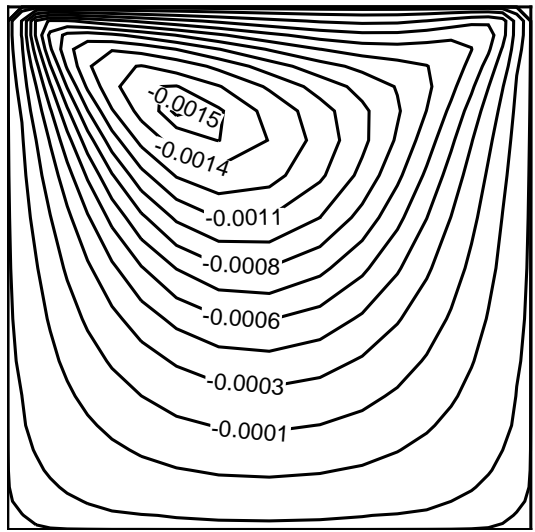
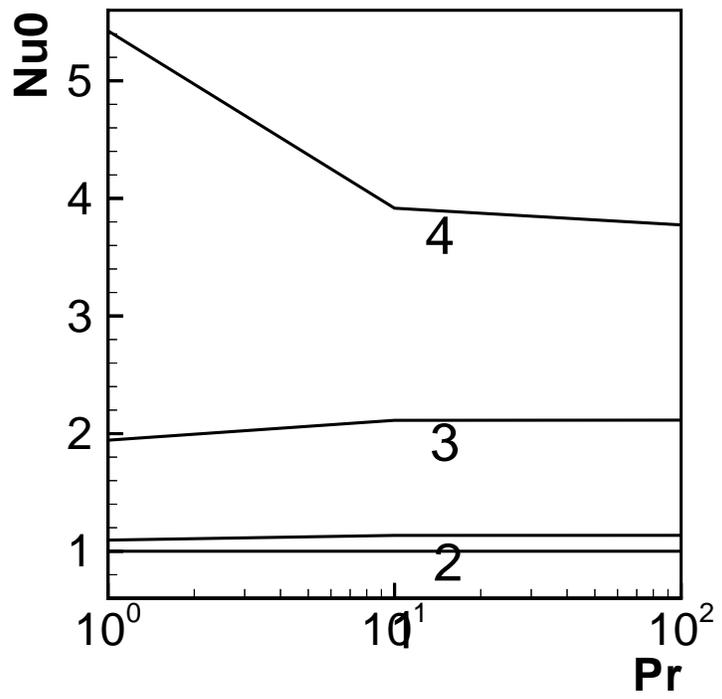
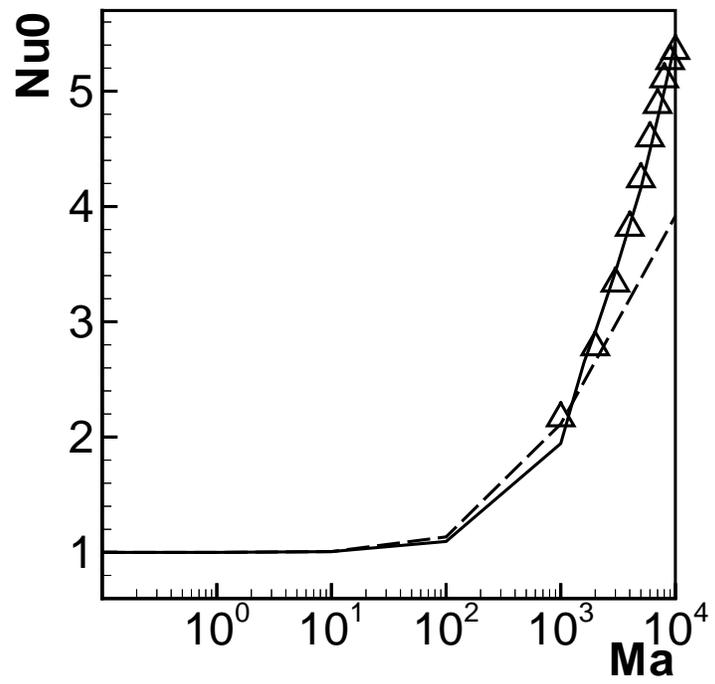
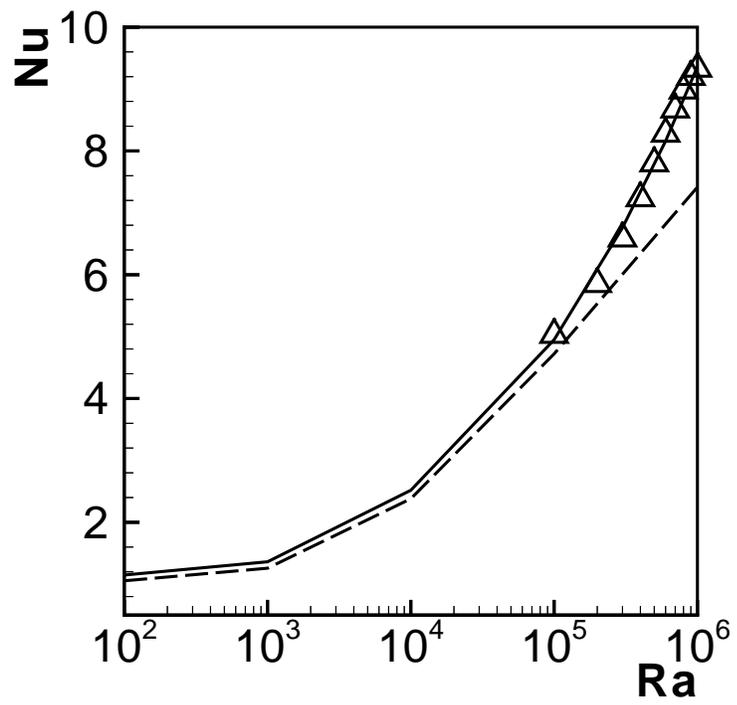
\end{document}